\documentclass[aps,prl,twocolumn,superscriptaddress]{revtex4}
\usepackage{graphicx,amsmath,amssymb}
\usepackage{epsfig}
\usepackage{xcolor}
\DeclareMathOperator{\sgn}{sgn}

\begin{document}

\title{SU(4) Symmetry Breaking Revealed by Magneto-optical Spectroscopy  in
Epitaxial Graphene }


\author{Liang Z. Tan}
\affiliation{Department of Physics, University of California at
Berkeley, and Materials Sciences Division, Lawrence Berkeley National
Laboratory, Berkeley, CA 94720}

\author{Milan Orlita}
\affiliation{Laboratoire National des Champs Magn$\acute{e}$tiques
Intenses, CNRS-UJF-UPS-INSA, B.P. 166, 38042 Grenoble Cedex 9,
France} \affiliation  {Institute of Physics, Charles University,
Ke Karlovu 5, CZ-121 16 Praha 2, Czech Republic}

\author{Marek Potemski}
\affiliation{Laboratoire National des Champs Magn$\acute{e}$tiques
Intenses, CNRS-UJF-UPS-INSA, B.P. 166, 38042 Grenoble Cedex 9,
France}

\author{James Palmer}
\affiliation{School of Physics, Georgia Institute of Technology,
Atlanta, Georgia 30332, USA}

\author{Claire Berger}
\affiliation{School of Physics, Georgia Institute of Technology,
Atlanta, Georgia 30332, USA} \affiliation{Institut N$\acute{e}$el,
CNRS-UJF B.P. 166, 38042 Grenoble Cedex 9, France}

\author{Walter A. de Heer}
\affiliation{School of Physics, Georgia Institute of Technology,
Atlanta, Georgia 30332, USA}

\author{Steven G. Louie$^*$}
\affiliation{Department of Physics, University of California at
Berkeley, and Materials Sciences Division, Lawrence Berkeley National
Laboratory, Berkeley, CA 94720}

\author{G\'{e}rard Martinez}
\affiliation{Laboratoire National des Champs Magn$\acute{e}$tiques
Intenses, CNRS-UJF-UPS-INSA, B.P. 166, 38042 Grenoble Cedex 9,
France}

\date{\today}
\begin{abstract}
Refined infra-red magneto-transmission experiments have been
performed in magnetic fields $B$ up to 35~T on a series of
multi-layer epitaxial graphene samples. Following the main optical
transition involving the $n$ = 0 Landau level, we observe a new
absorption transition increasing in intensity with magnetic fields
$B\ge 26$T. Our analysis shows that this is a signature of the
breaking of the SU(4) symmetry of the $n$ = 0 LL. Using a
quantitative model, we show that the only symmetry breaking scheme
consistent with our experiments is a charge density wave (CDW).
\end{abstract}


\maketitle

\section{Introduction}

In multicomponent quantum Hall systems, interaction effects
lead to a rich variety of broken symmetry ground states.
In graphene, the spin and valley degrees of freedom of the lowest Landau
level (LL) form an SU(4) symmetric quartet. Refined transport
experiments have shown evidence of a broken symmetry state
\cite{Zhang,Song,Miller,Zhao,Young,Yu,Amet,Young2},
but there is no clear consensus on its nature
\cite{Kharitonov,Abanin,Sodemann,Roy},
and how it is affected by different substrates and
disorder. While the spin degree of freedom has been probed in
tilted magnetic fields \cite{Zhang,Zhao,Young,Young2},
we show here that the valley degree of freedom can be accessed by
examining the signatures of optical phonons in magneto-transmission
spectra. In this paper, we show that our observation of a new
absorption transition supports the existence of a charge density wave
(CDW) in our epitaxial graphene samples.

In a quasiparticle picture, charge carriers in graphene are
characterized by a Dirac-like spectrum around the $K$ and $K'$
equivalent points (``valley'') of the Brillouin zone of the
hexagonal crystal lattice. As a consequence, the application of a
magnetic field $B$ perpendicular to the plane of the structure
splits the electronic levels into Landau levels (LL) indexed by
$n$, with specific energies $E_{n}=\sgn(n)v_{F}\sqrt{2e\hbar B|n|}$
where $n$ are integers including 0 ($v_{F}$ being the Fermi
velocity). In this paper, we are concerned with how a broken
symmetry phase can be observed in infra-red magneto-optical
transitions involving the $n$ = 0 LL (i.e., transitions from
$n=-1$ to $n=0$ or from $n=0$ to $n=1$ equivalent to a cyclotron
resonance (CR) transition in the quantum limit) with an energy
$E_{01}=v_{F}\sqrt{2e\hbar B}$ \cite{Henriksen}. Our previous work
\cite{Orlita} has reported on the magnetic field dependence of
this transition revealing its interaction with the $K$-phonon.
Besides this specific interaction, we observed that the basic
broadening $\gamma_{01} (B)\propto\sqrt{B}$ of the transition had
an additional component proportional to $B$ in contrast to all
theoretical models \cite{Yang}. This could be already a sign of
the breaking of the valley degeneracy.

\begin{figure}
  \includegraphics[width=0.95\columnwidth]{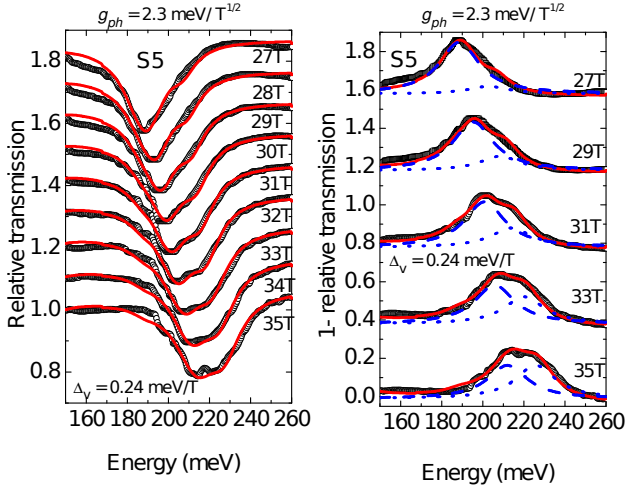}
  \caption{\label{figS5} (color online). Left panel: Evolution of the $E_{01}$ transition
  for different values of the magnetic field beyond 27~T.
  Experimental transmission data for sample S5 (open circles) is
  compared with calculated transmission spectra (red lines),
  for different values of magnetic field, using the proposed CDW model.
  Right panel: 1 - (relative transmission) measured in the experiment and
  calculated (red lines) for the CDW phase, for magnetic fields $B$ = 27, 29, 31,
  33, 35 T. Deconvolution of the experimental spectra into two
  Lorentzians is shown in blue dashed and dotted lines.
}
\end{figure}

In the present work, we use the $\Gamma$-phonon at the Brillouin
zone center as a probe of the valley symmetry breaking. In the
absence of valley symmetry breaking, the $\Gamma$-phonon does not
affect the infra-red absorption spectrum because the
electron-phonon matrix elements are of opposite signs for the $K$
and $K'$ valleys \cite{Goerbig}. However, one expects to see
signs of valley symmetry breaking when
the energy $E_{01}(B)$ is larger than that of the optical
$\Gamma$-phonon ($\hbar \omega_\Gamma = 0.196 \textrm{eV}$). It
turns out, indeed, that when that condition is reached, a new
optical transition develops at an energy \emph{higher} than the
main line (Fig.~\ref{figS5}). 
We interpret this as a signature of the breaking of the
SU(4) symmetry. A model has been established to reproduce these
findings and applied to the different phases which have been
proposed.

In Sec.~\ref{secExpt}, we discuss these experimental observations and methods in more detail. We first interpret our experimental findings within a simplified model for valley symmetry breaking in Sec.~\ref{secSimple} before deriving a more complete Hamiltonian in Sec.~\ref{secFull} and calculating the optical conductivity in various broken symmetry phases in Sec.~\ref{secPhases}. Finally, a comparison between experiment and theory is presented in Sec.~\ref{secCompare}, followed by conclusions in Sec.~\ref{secConc}.

\section{Experimental observations and methods \label{secExpt}}

\begin{figure}
\includegraphics [width=0.9\columnwidth]{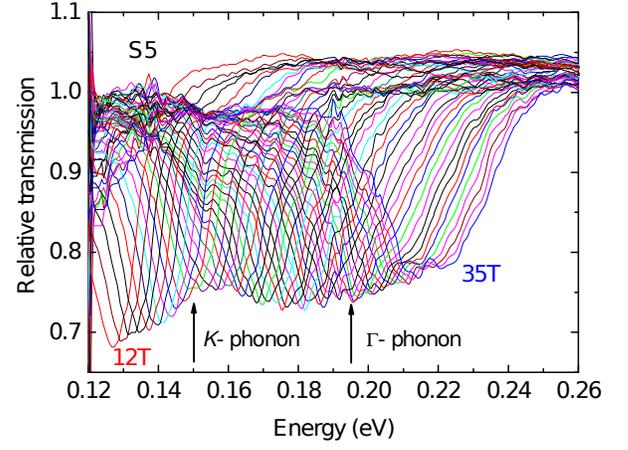}
\includegraphics [width=0.9\columnwidth]{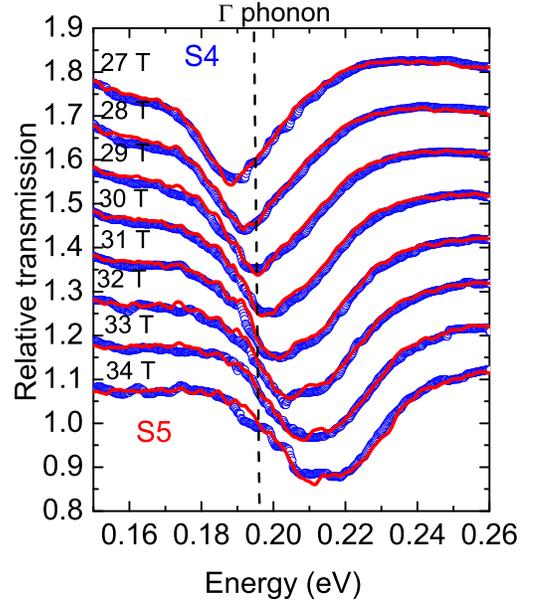}
\caption{\label{figS4S5}    (color online).
Top: Relative transmission spectra of
sample S5, for different magnetic field values up to 35~T.
Bottom: Evolution of the $E_{01}$ transition for
different values of the magnetic field between 27 T and 34 T, for
samples S4 (open dots) and S5 (full lines). }
\end{figure}

In our experiment, precise infra-red transmission measurements
were performed on multi-layer epitaxial graphene samples, at 1.8
K, under magnetic fields up to 35~T.
The light (provided and analyzed by a Fourier transform
spectrometer) was delivered to the sample by means of light-pipe
optics. All experiments were performed with nonpolarized light, in
the Faraday geometry with the wave vector of the incoming light
parallel to the magnetic field direction and perpendicular to the
plane of the samples. A Si bolometer was placed directly beneath
the sample to detect the transmitted radiation. The response of
this bolometer is strongly dependent on the magnetic field.
Therefore, in order to measure the absolute transmission TA
($B,\omega)$, we used a sample-rotating holder and measure for
each value of $B$ a reference spectrum through a hole. These
spectra are normalized in turn with respect to TA($0,\omega$) to
obtain a relative transmission spectrum TR($B,\omega$) which only
displays the magnetic field dependent features. Those spectra are
presented in Fig.~\ref{figS4S5}.

The samples were grown \cite{Berger} on the C-terminated surface of SiC
and display the characteristic transmission spectra of isolated graphene
monolayers that arise from rotational stacking of the sheets \cite{Hass}.
The thickness $d$
of the SiC substrate has to be reduced significantly in order to
minimize the very strong double-phonon absorption of SiC in the
energy range of interest. In the first series $d$ was reduced to
~$60 \mu m$ and related samples have been used to perform the
experiments reported earlier \cite{Orlita}. One of them, named S4,
was used to compare the data with those obtained on
sample S5 from a new series where the thickness $d$ was further
reduced down to ~$32 \mu m$. We compare in Fig.~\ref{figS4S5} the
transmission spectra, at high fields, for samples S4 and S5.
Technically speaking, the optical response of both samples is
almost the same, showing that they have a similar number of
active layers.

Taking into account all layer dielectric properties of each sample in a
multi-layer dielectric model, we determine the effective number, $N_{eff}$,
of graphene sheets with their respective carrier density
(see Appendix). For samples S4 and S5, we have found that
$N_{eff} = 7$ with carrier densities 
$\{5.5, 3.2, 1.8, 1.1, 0.5, 0.1, 0.1\} 10^{12} \textrm{cm}^{-2}$ and 
$\{6.0, 3.5, 2.0, 1.2, 0.5, 0.1, 0.1\} 10^{12} \textrm{cm}^{-2}$ respectively.
These carrier densities are fixed for each sample.

\begin{figure}
    \includegraphics [width=0.9\columnwidth]{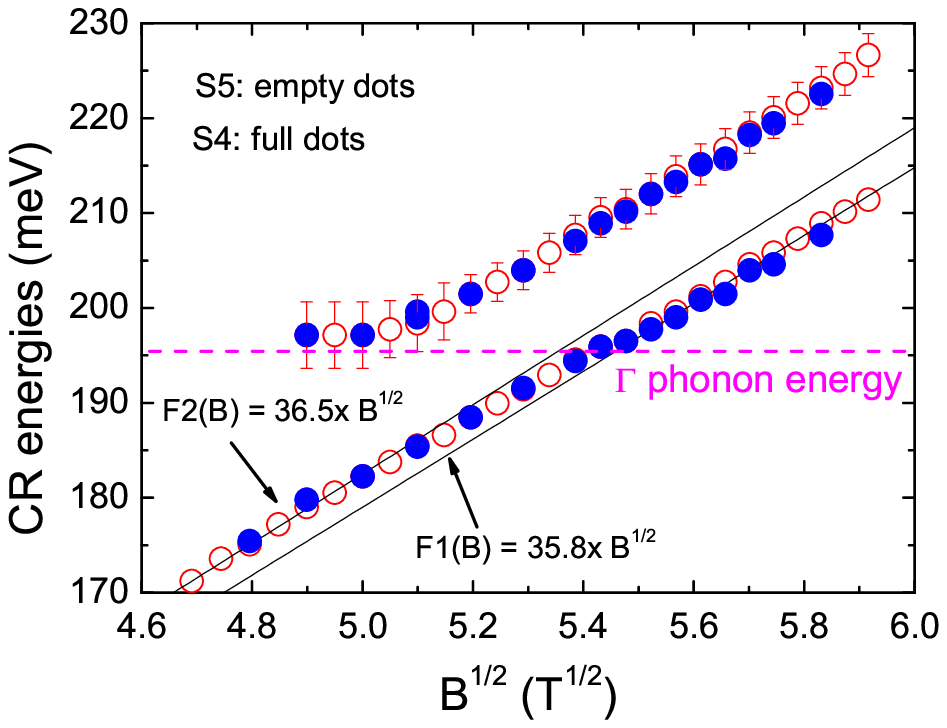}
    \includegraphics [width=0.9\columnwidth]{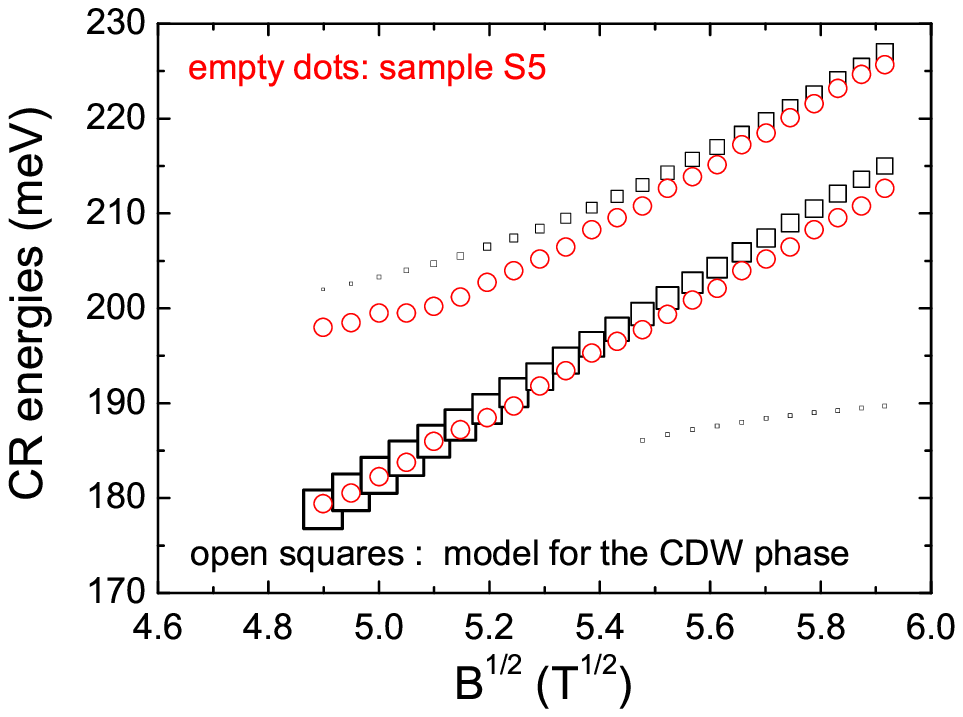}
\caption{\label{figEnCR} (color online). Top: CR energies, resulting
from the deconvolution of the experimental transmission traces,
as a function of $\sqrt{B}$ for samples S4 (full dots) and S5
(open dots). Error bars are similar for both samples. The full
lines are a linear fit of the data for the low energy transition
at low (F2($B$)) and high (F1($B$)) field.
  Bottom: Comparison of the variation of experimental CR energies for sample S5 (open dots),
  with that calculated for the CDW phase (open squares), as a function of $\sqrt{B}$.
  The size of the open squares mimics the relative oscillator strength of the optical transition. }
\end{figure}

The transmission spectra of sample S5 at high magnetic fields are displayed in
Fig.~\ref{figS5}.
We observe a new transition occurring at an
energy \emph{higher} than that of the CR line, growing in
intensity when increasing the magnetic field. This behavior cannot
be explained without breaking the  SU(4) symmetry in graphene. In
order to characterize more clearly these findings, one can treat
the data, as a first step and in a very rough way, extracting from
the transmission data the real part of the effective diagonal
component of the conductivity $\sigma_{xx}(\omega, B)$
\cite{Sadowski}. We have deconvoluted this result with two
Lorentzians of equal width, extracting the evolution of the two
extrema with the magnetic field. The resulting energies are
displayed in Fig.~\ref{figEnCR} (top panel) for samples S4 and S5.

Though the procedure adopted at this initial level is quite rough,
it provides important information: (i) The evolution of the lower
energy line varies at low fields like $B^{1/2}$ (function F2($B$)
in Fig.~\ref{figEnCR}) with a coefficient proportional to the Fermi
velocity $v_{F}$ and ends at higher fields with a similar
dependence (function F1($B$)) but with a smaller value of $v_{F}$
which is, by itself, a sign of some interaction occurring at an
energy close to that of the $\Gamma$-phonon; (ii) The second
component of the deconvolution always appears at energies larger
than that of the $\Gamma$-phonon; (iii) In principle, in the SU(4)
symmetric picture, it is not possible to explain the occurrence of
an additional transition, growing in intensity with $B$, at
\emph{higher} energies than the main transition line; (iv) It is
therefore clear that the $\Gamma$-phonon plays a crucial role
though it should not, indicative that the SU(4) symmetry is
broken. Using these observations, we now have some guidelines to
develop a theory which can explain quantitatively the experimental
observations. In addition, we note  that results for samples S4
and S5 are quite similar within the experimental errors. Knowing
that the active layers which contribute to the $E_{01}$ transition
should have a filling factor $\nu \leq 2$ ( $\nu= N_{s}\Phi_{0}/B,
\Phi_{0}$ being the flux quantum and $N_{s}$ the carrier density)
and that, in samples S4 and S5, the carrier
density for active layers do not have the same sequence, the
physical mechanisms describing the experimental findings should
not be very dependent on the doping of active layers. This is
indeed the case as discussed below.

\section{Simplified model for valley symmetry breaking\label{secSimple}}

To illustrate how the electron-phonon interaction and the valley
symmetry breaking give rise to the observed features in the
transmission spectrum,
we first introduce a simplified model for the interaction of the
$\Gamma$ phonon with the $E_{01}$ excitation, before discussing the
full SU(4) calculation. The simplified model provides a
minimal description of the valley symmetry breaking by neglecting the spin
degree of freedom in the $n=0$ LL. We assume that $K$ and $K'$ sublevels
of the $n=0$ LL are separated in energy by $\Delta_V$, and have
different filling factors $\nu_K$ and $\nu_{K'}$. Considering just the
$n=0$ to $n=1$ transitions, the interaction with
the $\Gamma$ phonon is captured by the Hamiltonian (in the basis of creating
an electronic excitation in $K$, electronic excitation in $K'$, and a $\Gamma$
phonon, in that order)

\begin{equation}\label{H}
H=
\begin{pmatrix}
  E_{01}-\Delta_{V}/2 & 0 & g_{ph}\sqrt{\nu_K}\\
                    0 & E_{01}+\Delta_{V}/2 & -g_{ph}\sqrt{\nu_{K'}}\\
        g_{ph}\sqrt{\nu_K} & -g_{ph}\sqrt{\nu_{K'}} & \hbar\omega_{\Gamma}
\end{pmatrix}.
\end{equation}

\noindent where $g_{ph}$ characterizes the
electron-$\Gamma$-phonon interaction. The optical conductivity is
calculated using the Green's function formalism introduced by
Toyozawa \cite{Toyozawa1}. The diagonal component of the conductivity is:

\begin{equation}\label{green}
  \sigma_{xx}(\hbar\omega) = \frac{1}{\omega} \textrm{Im} M^\dagger_{x}G(\hbar\omega)M_{x}
\end{equation}

\noindent where the Green's function is $G = \left(
\hbar\omega-H-i\eta \right)^{-1}$, with $\eta \rightarrow 0^+$ (see
Sec.~\ref{ssOp}).
The optical matrix elements for the simplified model are
$M_x^\dagger = \left(\sqrt{\nu_K}, \sqrt{\nu_{K'}},0 \right) $.
The simplified model explains the splitting of the main transition
line in the two limits $E_{01}\approx\hbar \omega_\Gamma$ and
$E_{01} \gg \hbar \omega_\Gamma$. In the absence of valley
splitting ($\Delta_V=0$ and $\nu_K = \nu_{K'}$), the eigenstates
of the pure electronic part of $H$ in 
Eq.~\ref{H} are valley-symmetric and valley-antisymmetric
combinations of $E_{01}$ transitions, i.e. $\frac{1}{2}\left(
c^\dagger_{1,K}c_{0,K} + c^\dagger_{1,K'}c_{0,K'} \right) \lvert
\textrm{GS} \rangle$ and $\frac{1}{2}\left( c^\dagger_{1,K}c_{0,K}
- c^\dagger_{1,K'}c_{0,K'} \right) \lvert \textrm{GS} \rangle$,
respectively, where $\rvert{}\textrm{GS}\rangle$ denotes the ground
state and $c^\dagger_{n,K}$ are creation operators at LL $n$ and
valley $K$. The valley-symmetric combination is infra-red active
but does not interact with the $\Gamma$ phonon, while the
valley-antisymmetric combination is infra-red inactive and
interacts with the $\Gamma$ phonon. The symmetry breaking valley
splitting term $\Delta_V$ allows both eigenmodes to interact with
the $\Gamma$ phonon while remaining infra-red active, inducing a
splitting of the main transmission line in the vicinity of the
$\Gamma$ phonon frequency. Away from the $\Gamma$ phonon frequency
($E_{01} \gg \hbar \omega_\Gamma$), $E_{01}$ transitions at $K$
and $K'$ interact weakly with the phonon and the splitting of the
main transmission line is controlled directly by the energy
difference $\Delta_V$.

\section{Theory of magneto-phonon resonance in the presence of SU(4) symmetry breaking \label{secFull}}

We reintroduce the spin degree of freedom and the $n=-1$ to $n=0$
transitions in order to obtain a quantitative understanding of the
experiment. We consider different theoretical models of the $n=0$
LL SU(4) symmetry breaking, taking into account the effects of
$\nu\ne0$ and disorder  by introducing Gaussian broadening into a
mean field theory (Sec.~\ref{ssGS}). Different
symmetry-breaking phases are represented in the mean field theory
by different orderings and filling factors of the four sublevels
of the $n=0$ LL. We consider four candidate symmetry-breaking
phases that have been proposed in the literature
\cite{Kharitonov}: Ferromagnetic(F), Charge Density Wave (CDW),
Canted Antiferromagnetic (CAF) and Kekul\'{e}-distortion (KD), and
calculate the optical conductivity using Eq.~\ref{green} with the
appropriate Hamiltonian $H$ for each phase.
Treating these phases on the same footing (detailed in
Sec.~\ref{secPhases}), we find that each phase results in
characteristic features in the evolution of the transmission
spectrum as a function of the magnetic field. By examining the
intensities and positions of the transmission lines, we identify
the symmetry broken phase in the samples used in our experiment as
the CDW type \cite{Fuchs,Jung}.

\subsection{Description of the ground state \label{ssGS}}

We assume that the ground state is a single Slater determinant of
the form:
\begin{equation}\label{GS}
|GS\rangle = \prod_{j=1}^{4}
\prod_{m_{j}=1}^{N_{j}}\Psi_{j,m_{j}}^{\dagger} |0\rangle
\end{equation}

\noindent the index \emph{j} runs over the 4-dimensional spin/valley space
and $m_{j}$ describe the "guiding center" degree of freedom. The
state ($j,m_{j}$) is represented by the wavefunction
$\xi_{j}\phi_{m_{j}}(\overrightarrow{r})$ where $\xi_{j}$ is a
four-component spinor and $\phi_{m_{j}}(\overrightarrow{r})$ is
the orbital part of the wavefunction. These wavefunctions belong
to the $n=0$ landau level (LL) of graphene. The occupation numbers
$N_{j}$ count the number of $j$ states that are occupied in this
ground state.

There are different models proposed to describe the
symmetry-broken phase of graphene which have been reviewed by
Kharitonov \cite{Kharitonov}. For a given model, we assume that
the system is polarized along a certain direction in $j$-space.
For instance, with increasing order of energies, $j=1,2,3,4$
corresponds to ($K'\uparrow, K'\downarrow, K\uparrow,
K\downarrow$) in the charge density wave (CDW) phase. The remaining
degrees of freedom, $\phi_{m_{j}}(\overrightarrow{r})$ and
$N_{j}$, are treated as variational parameters,
subject to the constraint $N_1+N_2+N_3+N_4 = N$. We minimize the
energy of the ground state $ E_{GS}=\langle GS|H_{0}+
H_{e-e}+H_{disorder}|GS\rangle$. Here $H_{0}$ is the single part
of the Hamiltonian without disorder, $H_{e-e}$ the interaction
term and $H_{disorder}$ the disorder potential. Because we assume
a single Slater determinant, we can apply mean-field theory and
obtain single-particle energy levels $E_{j,m_{j}}$ (The origin of the
energies is taken to be at the energy of the $n=0$ LL of the
non-interacting system).

In a system with finite disorder, the energy levels $E_{j,m_{j}}$ are
clustered about mean values $E_j=\underset{m_j}{\textrm{avg }} E_{j,m_{j}}$.
We remove the $m_{j}$ degrees of freedom by replacing the energy
levels $E_{j,m_{j}}$ by broadened energy levels centered at $E_j$.
There is a Fermi level $E_{F}$ which fixes the occupation numbers
$N_{j}$ when the graphene layer is doped with a total filling factor
$\nu$. Assuming the broadening to be of
Gaussian type with a width $\gamma_{0}$ the Fermi level is
determined by solving the following equation:

\begin{equation}\label{EF}
\nu = \sum_{j}\textrm{Erf}(\frac{E_{F}-E_{j}}{\sqrt{2}\gamma_{0}})
\end{equation}

\noindent from which one can calculate the individual filling factors
$\nu_{j}=(1+ \textrm{Erf}(\frac{E_{F}-E_{j}}{\sqrt{2}\gamma_{0}}))/2$ for
each level $E_{j}$. These $E_{j}$ will be used, later on, as
fitting parameters dependent on the broken-symmetry phase under
consideration. Note that in this approach \emph{all}
optical transitions to or from the $n=0$ LL are allowed.

\subsection{Description of the optical transitions \label{ssOp}}

We first consider the transitions from the $n=0$ LL to $n=1$ LL.
The Hamiltonian of the magneto-excitons, including their
interaction with the $\Gamma$-phonon, denoted
$H_{\circlearrowright}$ (reminding that it describes the optical
transitions allowed in the $\sigma^{+}$ polarization), is:

\begin{widetext}
\begin{equation}\label{H+}
H_{\circlearrowright} =
\begin{pmatrix}
\hbar\omega_{01}-E_{1} & 0 & 0 & 0 & g_{1}\sqrt{\nu_{1}}  \\
0 & \hbar\omega_{01}-E_{2} & 0 & 0 & g_{2}\sqrt{\nu_{2}} \\
0 & 0 & \hbar\omega_{01}-E_{3} & 0 & g_{3}\sqrt{\nu_{3}} \\
0 & 0 & 0 & \hbar\omega_{01}-E_{4} & g_{4}\sqrt{\nu_{4}}  \\
g_{1}^{*}\sqrt{\nu_{1}} & g_{2}^{*}\sqrt{\nu_{2}} & g_{3}^{*}\sqrt{\nu_{3}} & g_{4}^{*}\sqrt{\nu_{4}} &  \hbar\omega_{ph}  \\
\end{pmatrix}.
\end{equation}
\end{widetext}

\noindent where $\hbar\omega_{01}$ is the energy of the $E_{01}$ transition
from the $n=0$ LL to $n=1$ LL in the absence of interactions and
$\hbar\omega_{ph} $ that of the $\Gamma$-phonon. This Hamiltonian
describes the excitations from the 4 sublevels $\{E_j, j=1..4\}$ of
the $n=0$ LL to the $n=1$ LL. The matrix elements $\{g_j, j=1..4\}$
respectively describe their interaction with the $\Gamma$-phonon, and is
dependent on the wavefunction character of the 4 sublevels (i.e.,
dependent on the broken-symmetry phase).
In general, $g_{j} \propto \sqrt{B}$ \cite{Goerbig}, with a
prefactor dependent on $j$ and the broken-symmetry phase.

Similarly, the Hamiltonian describing the magneto-excitons for the
transitions from  the $n=-1$ LL to  the $n=0$ LL (allowed in the
$\sigma^{-}$ polarization) is written as:

\begin{widetext}
\begin{equation}\label{H-}
H_{\circlearrowleft} =
\begin{pmatrix}
\hbar\omega_{01}+ E_{1} & 0 & 0 & 0 & g_{1}\sqrt{1-\nu_{1}}  \\
0 & \hbar\omega_{01}+ E_{2} & 0 & 0 & g_{2}\sqrt{1-\nu_{2}} \\
0 & 0 & \hbar\omega_{01}+ E_{3} & 0 & g_{3}\sqrt{1-\nu_{3}} \\
0 & 0 & 0 & \hbar\omega_{01}+ E_{4} & g_{4}\sqrt{1-\nu_{4}}  \\
g_{1}^{*}\sqrt{1-\nu_{1}} & g_{2}^{*}\sqrt{1-\nu_{2}} & g_{3}^{*}\sqrt{1-\nu_{3}} & g_{4}^{*}\sqrt{1-\nu_{4}} &  \hbar\omega_{ph}  \\
\end{pmatrix}.
\end{equation}
\end{widetext}

\noindent The total Hamiltonian $H$ describing the magneto-excitons is
therefore:
\begin{equation}\label{Htot}
H =
\begin{pmatrix}
H_{\circlearrowright} & 0   \\
0 & H_{\circlearrowleft}  \\
\end{pmatrix}.
\end{equation}

We will also need to
introduce the optical matrix elements $M_{x}$ and $M_{y}$ for
corresponding transitions. These matrix elements depend on the ground
state under consideration. 
In CDW case they are (see Sec.~\ref{ssCDW}) : 
$M_{x}/v_{0}=\{\sqrt{\nu_{1}}$, $\sqrt{\nu_{2}}$, $\sqrt{\nu_{3}}$,
$\sqrt{\nu_{4}}$, 0, $-\sqrt{1-\nu_{1}}$, $-\sqrt{1-\nu_{2}}$,
$-\sqrt{1-\nu_{3}}$, $-\sqrt{1-\nu_{4}}$, $0\}$
and $M_{y}= i M_{x}$. For a different scenario, 
the optical matrix elements will be transformed to a different basis,
as will be detailed in Sec.~\ref{secPhases}.

The Green's function for the magneto-excitons, is obtained as $G =
((\hbar\omega+i\gamma_{01}).\text{I}- H)^{-1}$ (where I is the unit matrix and  $\gamma_{01}$ the broadening of the $E_{01}$ transition).
This allows us to calculate the different components of the conductivity:

\begin{equation}\label{Sigma}
\begin{aligned}
\sigma_{xx}(\omega)= \frac{i}{\omega} M_{x}^{T}.G.M_{x}\\
\sigma_{xy}(\omega)= \frac{1}{\omega}M_{x}^{T}.G.M_{y}^{*}
\end{aligned}
\end{equation}

\section{Optical conductivity in the different phases \label{secPhases}}

Here, we calculate the optical conductivity for the different symmetry broken phases, using Eq.~\ref{Sigma}. 

\subsection{Charge density wave (CDW) phase \label{ssCDW}}

The CDW phase is characterized, at filling factor $\nu=0$,
by two electronic LL full in one valley (say $K'$ for instance) and two
LL empty in the other valley ($K$).
We will introduce a valley asymmetry $\Delta_V$ mainly
determined by electron-electron interactions \cite{Kharitonov} and
a Zeeman splitting $\Delta_S$. Therefore the sequence of sublevels
take the following form:

\begin{equation}
  \begin{aligned}
    E_{1}(K' \uparrow) & = -\Delta_V/2- \Delta_S/2 \\
    E_{2}(K' \downarrow) & = -\Delta_V/2+ \Delta_S/2 \\
    E_{3}(K \uparrow) & = \Delta_V/2- \Delta_S/2 \\
    E_{4}(K \downarrow) & = \Delta_V/2+ \Delta_S/2
  \end{aligned}
\end{equation}

In this case, the parameters governing the
electron-$\Gamma$phonon interaction $g_{1}$, $g_{2}$ on one hand
and $g_{3}$, $g_{4}$ on the other hand are of opposite sign. That is,

\begin{widetext}
\begin{equation}\label{cdweph}
  \begin{aligned}
    \langle GS_{CDW} + \Gamma\textrm{phonon} \rvert H_{e-ph}
    \Psi^\dagger_{Ks,1}
    \Psi_{Ks,0} \lvert GS_{CDW} \rangle & = g_{ph} /\sqrt{2} \\
    \langle GS_{CDW} + \Gamma\textrm{phonon} \rvert H_{e-ph}
    \Psi^\dagger_{K's,1}
    \Psi_{K's,0} \lvert GS_{CDW} \rangle & = -g_{ph} /\sqrt{2} \\
  \end{aligned}
\end{equation}
\end{widetext}

\begin{figure}
\includegraphics [width=0.95\columnwidth]{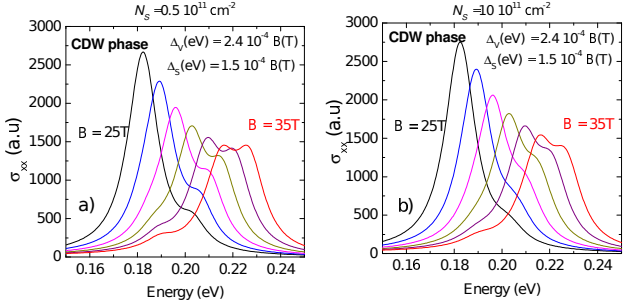}
\caption{\label{figCDW} (color online). CDW phase with $\Delta_V\propto B$:
evolution of the $\sigma_{xx}$ component of the conductivity  for
different values of the magnetic field between 25 T and 35 T  for
a) a carrier density $N_{s}= 0.5 \times 10^{11} \text{cm}^{-2}$
and b) $N_{s}= 1\times 10^{12} \text{cm}^{-2}$. In both cases
$g_{ph}= 2.3 \times\sqrt{B[T]}$ meV. }
\end{figure}

\noindent where $\Psi^\dagger_{K,s,n}$ is the creation operator for 
electrons in valley $K$, spin $s$, Landau level $n$. On the other hand,
the electron-light interaction (which determines $M$)
has the same sign at both valleys.

\begin{equation}\label{cdwelight}
  \begin{aligned}
    \langle GS_{CDW} + \textrm{photon} \rvert H_{e-light}
    \Psi^\dagger_{Ks,1}
    \Psi_{Ks,0} \lvert GS_{CDW} \rangle & = 1 \\
    \langle GS_{CDW} + \textrm{photon} \rvert H_{e-light}
    \Psi^\dagger_{K's,1}
    \Psi_{K's,0} \lvert GS_{CDW} \rangle & = 1 \\
  \end{aligned}
\end{equation}

\noindent The results obtained for this phase are presented in
Fig.~\ref{figCDW}, assuming $\Delta_V$  proportional to $B$,
for two extreme values of the carrier density. The electron-phonon coupling was taken to be $g_{ph}= 2.3 \times\sqrt{B[T]}$ meV, which agrees
with density functional theory (DFT) calculations \cite{Piscanec}
and experiments \cite{Lazzeri,Yan,Pisana}.
The results are not very dependent on $N_s$.
The value of $\Delta_S= 0.15$ meV $B$ corresponds to a
g-factor of 2.6 to be compared with $2.7\pm0.2$ reported in
\cite{Kurganova}. The splitting of the
transition is directly governed by the amplitude of
$\Delta_V$ whereas the introduction of $\Delta_S$
modifies only the relative amplitude of the two transitions. In
all cases \textit{both} $\Delta_V$ and $g_{ph}$ need to be
finite to observe the effect. We finally note that, in this case,
$\Delta_V > \Delta_S$ in coherence with the assumption
made in our previous work \cite{Orlita}.

\begin{figure}[h]
\includegraphics [width=0.95\columnwidth]{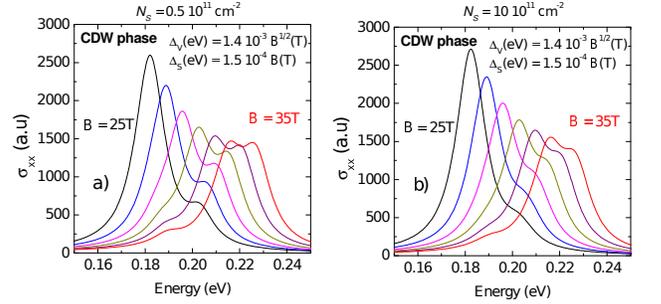}
\caption{\label{figCDW2} (color online). CDW phase with $\Delta_V\propto
\sqrt{B}$: evolution of the $\sigma_{xx}$ component of the
conductivity  for different values of the magnetic field between
25 T and 35 T  for a) a carrier density $N_{s}= 0.5 \times 10^{11}
\text{cm}^{-2}$ and b) $N_{s}= 1\times 10^{12} \text{cm}^{-2}$. In
both cases $g_{ph}= 2.3 \times\sqrt{B[T]}$ meV. }
\end{figure}

However there is no clear consensus about the field dependence on
$\Delta_V$ \cite{Kharitonov}. Therefore one can
alternatively assume that $\Delta_V$ is proportional to
$\sqrt{B}$. The corresponding results are displayed in
Fig.~\ref{figCDW2} keeping all other parameters fixed. We obtained
essentially the same results as in Fig.~\ref{figCDW}. Within the
experimental errors we will not be able to differentiate between
the two magnetic field variations of $\Delta_V$.

The CDW state is compatible with the experimental results
as we will see below.

\subsection{Kekul\'{e}-distortion  (KD) phase \label{ssKD}}

In this phase \cite{Kharitonov}, the $K$ and $K'$ valleys
hybridize into linear combinations $\overline{K}$,
$\overline{K'}$. At $\nu=0$, both spin $\uparrow$ and spin $\downarrow$
electrons occupy one of these valley-combinations, say
$\overline{K}$. The $\nu=0$ ground state for the KD phase is
$\Psi_{\overline{K}\uparrow, 0}^{\dagger}\Psi_{\overline{K}\downarrow,
0}^{\dagger}|0\rangle$. Therefore, the "natural" basis for this phase,
where the density matrix is diagonal, is $\{\overline{K}\uparrow,
\overline{K}\downarrow, \overline{K'}\uparrow,
\overline{K'}\downarrow\}$ in contrast to the basis $\{K\uparrow,
K\downarrow, K'\uparrow, K'\downarrow\}$ used in the CDW phase.
Therefore the sequence of sublevels take the following form:

\begin{equation}
  \begin{aligned}
    j=1 & : \overline{K} \uparrow \\
    j=2 & : \overline{K} \downarrow \\
    j=3 & : \overline{K'} \uparrow \\
    j=4 & : \overline{K'} \downarrow
  \end{aligned}
\end{equation}

\noindent The transformation rules for the operators in this basis are

\begin{equation}\label{changebasis}
  \begin{aligned}
    \Psi_{\overline{K},s,n}^\dagger = & \frac{1}{\sqrt{2}} 
      \left ( \Psi_{K,s,n}^\dagger + e^{i\phi} \Psi_{K',s,n}^\dagger \right ) \\
    \Psi_{\overline{K'},s,n}^\dagger = & \frac{1}{\sqrt{2}} 
      \left ( \Psi_{K,s,n}^\dagger - e^{i\phi} \Psi_{K',s,n}^\dagger \right )
  \end{aligned}
\end{equation}

\noindent where $\Psi^\dagger_{\overline{K},s,n}$ is the creation operator for 
electrons in valley state $\overline{K}$, spin $s$, Landau level $n$.
Making use of this change of basis (Eq.~\ref{changebasis}) and
Eq.~\ref{cdweph} and Eq.~\ref{cdwelight}, 
we derive that the electron-light matrix
elements do not change with respect to the CDW phase, and the
electron-phonon matrix elements $g$ vanish by symmetry. That is,

\begin{equation}
  \begin{aligned}
    \langle GS_{KD} + \Gamma\textrm{phonon} \rvert H_{e-ph}
    \Psi^\dagger_{\overline{K}\uparrow,1}
    \Psi_{\overline{K}\uparrow,0} \lvert GS_{KD} \rangle & = 0 \\
    \langle GS_{KD} + \textrm{photon} \rvert H_{e-light}
    \Psi^\dagger_{\overline{K}\uparrow,1}
    \Psi_{\overline{K}\uparrow,0} \lvert GS_{KD} \rangle & = 1
  \end{aligned}
\end{equation}

\noindent and the same for $\overline{K'}$. The
structure of the Hamiltonian (Eq.\ref{Htot}) becomes only diagonal
and no splitting is observed when calculating the conductivity.
Therefore the KD phase does not explain the experimental results.

\subsection{Ferromagnetic (F) phase \label{ssF}}

\begin{figure}[h]
\includegraphics [width=0.9\columnwidth]{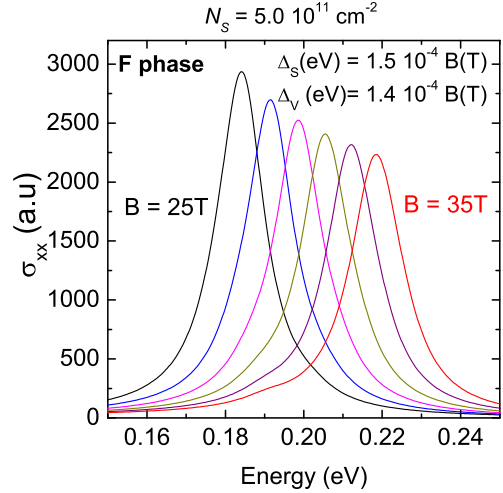}
\caption{\label{figF} (color online). Ferromagnetic phase with
$\Delta_V\propto B$: evolution of the $\sigma_{xx}$
component of the conductivity  for different values of the
magnetic field between 25 T and 35 T  for  a carrier density
$N_{s}= 5.0 \times 10^{11} \text{cm}^{-2}$  and  $g_{ph}= 2.3$
meV/$\text{T}^{1/2}$. }
\end{figure}

In the F phase \cite{Kharitonov}, the ground state, at filling
factor $\nu=0$, is composed in
both valleys $K$ and $K'$ of a single full LL with the same spin.
In analogy with the CDW phase, we will introduce a valley
asymmetry $\Delta_V$ and a Zeeman splitting $\Delta_S$. Therefore
the sequence of energy levels take the following form:

\begin{equation}
  \begin{aligned}
    E_{1}(K' \downarrow) & = -\Delta_V/2- \Delta_S/2 \\
    E_{2}(K \downarrow) & = \Delta_V/2 - \Delta_S/2 \\
    E_{3}(K' \uparrow) & = -\Delta_V/2+ \Delta_S/2 \\
    E_{4}(K \uparrow) & = \Delta_V/2+ \Delta_S/2
  \end{aligned}
\end{equation}

Note that, in this case, $\Delta_S$ should
be larger than $\Delta_V$ to preserve the ferromagnetic
nature of the state. In the present case the parameters governing
the electron-$\Gamma$phonon interaction (Eq.\ref{H+},\ref{H-})
$g_{1}$, $g_{3}$ on one hand and $g_{2}$, $g_{4}$ in the other
hand are of opposite sign.

The results are displayed in Fig.~\ref{figF} where we have taken
for $\Delta_S$ the same evolution that in the CDW phase and
$\Delta_V\propto B$. The conductivity does not show any
significant splitting of the main line. In fact there is an
eigenvalue of the corresponding Hamiltonian larger than that of
the main line but it remains optically inactive. Therefore here
also, the F phase does not explain the experimental results.

\subsection{Canted anti-ferromagnetic (CAF) phase \label{ssCAF}}

The CAF phase for the ground state is described by a spin in
direction $\theta_{K}$ in valley $K$ and a spin in direction
$\theta_{K'}$ in valley $K'$. (The directions $\theta_{K}$ and
$\theta_{K'}$ are in general not opposite to each other except in
the special case of the anti-ferromagnetic phase). The direction
$\theta_{K}$ is oriented at an angle $\theta$ relative to the
magnetic field $B$ and the direction $\theta_{K'}$ at an angle
$-\theta$ with respect to it. (In the anti-ferromagnetic phase,
$\theta=\pi/2$). Here the Zeeman splitting should vary like
$\Delta_S\propto \cos\theta$ and if $\theta$ is close to $\pi/2$
this term should not play a dominant role. We choose the following
order of states:

\begin{equation}
  \begin{aligned}
    E_{1} (K,\theta_{K}) & = -\Delta_1/2- \Delta_2/2 \\
    E_{2} (K',\theta_{K'})& = -\Delta_1/2 + \Delta_2/2 \\
    E_{3} (K,\pi+\theta_{K}) & = \Delta_1/2- \Delta_2/2 \\
    E_{4} (K',\pi+\theta_{K'})& = \Delta_1/2 + \Delta_2/2
  \end{aligned}
\end{equation}

\noindent where the introduction of $\Delta_1$ reflects the CAF
pattern of spin. We assume in addition that the asymmetry between
valleys is reflected by  $\Delta_2$ (favoring here the $K$ valley).
To preserve the CAF phase $\Delta_2$ should be smaller than
$\Delta_1$. Similar to the F phase,  the parameters governing the
electron-$\Gamma$phonon interaction (Eq.\ref{H+},\ref{H-})
$g_{1}$, $g_{3}$ on one hand and $g_{2}$, $g_{4}$ in the other
hand are of opposite sign.

\begin{figure}[h]
\includegraphics [width=0.95\columnwidth]{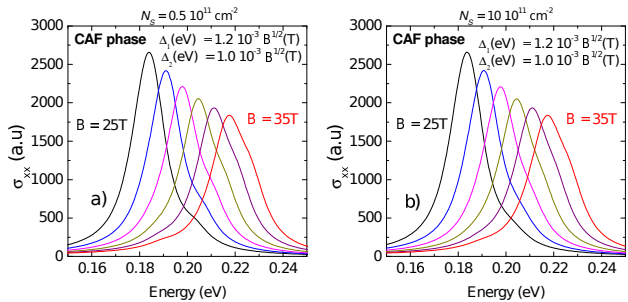}
\caption{\label{figCAF} (color online). CAF phase with $\Delta_1\propto \sqrt{B}$
and $\Delta_2\propto \sqrt{B}$: evolution of the $\sigma_{xx}$
component of the conductivity for different values of the magnetic
field between 25 T and 35 T for a) a carrier density $N_{s}= 0.5
\times 10^{11} \text{cm}^{-2}$ and b) $N_{s}= 1\times 10^{12}
\text{cm}^{-2}$. In both cases $g_{ph}= 2.3$ meV/$\text{T}^{1/2}$. }
\end{figure}

The results are displayed in Fig.~\ref{figCAF} where we have taken,
 $\Delta_1$ and $\Delta_2$ proportional to $\sqrt{B}$. The results are not very dependent on
the carrier concentration. We observe indeed a splitting of the
transition when \emph{both } $\Delta_2$ and $g_{ph}$ are different
from zero : in fact the splitting is governed by $\Delta_2$. In the
present case we do not have, \textit{a priori}, a guide for
choosing the values of $\Delta_1$ and $\Delta_2$. In order to be
consistent with experimental results, we have taken for $\Delta_1$
a value which provides an upper transition energy close to that
observed.

However the evolution of the spectra does not reflect the
experimental observations:  whatever is the choice of parameters,
the intensity of the high energy transition never reaches that of
the main transition in contrast to the CDW phase where it should
become dominant at fields higher than 35 T. This is discussed further in the next section.

\section{Comparison of experiment and theory \label{secCompare}}

\begin{figure}
  \includegraphics[width=0.95\columnwidth]{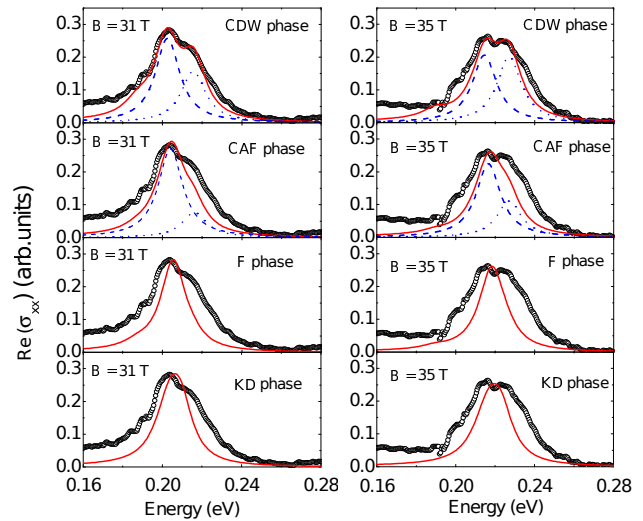}
  \caption{\label{figCandidates} (color online).
    Real part of the conductivity measured in the experiment (open
    dots) and calculated with the CDW, CAF, F and KD phases (red continuous lines), for magnetic
    fields $B= 31 T$ (left panel) and  $35 T$ (right panel).
    The Deconvolution of the model into two Lorentzians is shown for the CDW and CAF phases (blue dashed and dotted lines).
  }
\end{figure}

The KD, F and CAF phases result in transmission spectra
incompatible with experiment (Fig.~\ref{figCandidates}). In the KD phase,
electrons occupy linear combinations of the $K$ and $K'$ valleys;
the electron-phonon matrix elements vanish by symmetry, resulting
in a single transmission line. For the F phase, the occupancy of
the $K$ and $K'$ valleys are almost equal (Sec.~\ref{ssF}), and there is no significant splitting of the main
transmission line (Fig.~\ref{figCandidates}) . Similarly, the calculated
transmission spectra for the CAF phase   show a second CR line of
much lower intensity than the main CR line. Deconvoluting these
spectra with two Lorentzians we find a ratio of the CR weights of
$0.9\pm 0.05$ for the experiment, to be compared to the value 0.9
for the CDW phase and 0.4 for the CAF phase. Despite the
introduction of valley asymmetry into the CAF phase, we find that
it cannot explain the observed evolution of CR energies in the
experiment.

The CDW phase has unequal occupation
numbers of the $n=0$ LL at the $K$ and $K'$ valleys, corresponding
to a density modulation of the graphene $A$ and $B$ sublattices in
real space. Unlike the ideal disorder-free CDW discussed in
\cite{Kharitonov}, both $K$ and $K'$ valleys have non-zero
occupation number in our calculation, due to disorder-induced
broadening. Nevertheless, the mechanism giving
rise to the splitting of the $E_{01}$ transmission line remains
essentially the same as illustrated by the simple model
(Eq.~\ref{H}) above.

\begin{figure}[h]
\includegraphics [width=0.95\columnwidth]{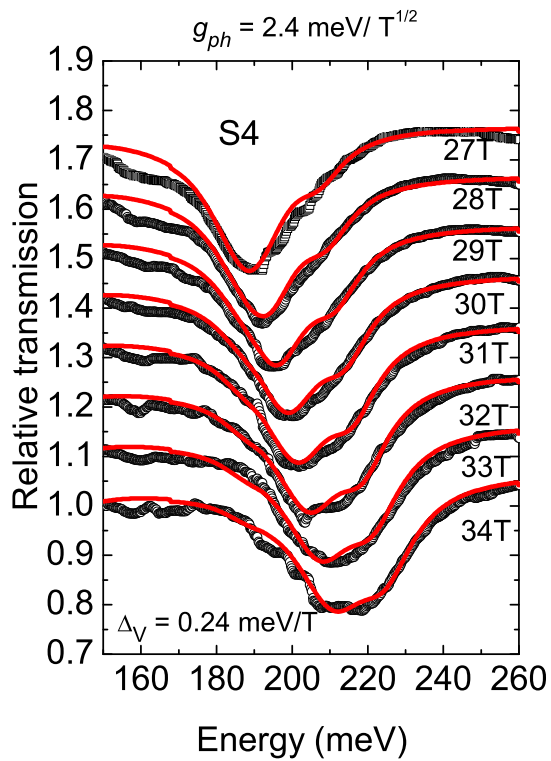}
\caption{\label{figS4} (color online). Fit of the experimental data (open dots)
for sample S4 with the conductivity model derived for the CDW
phase (continuous lines) using $\Delta_V=0.24 $ meV/T and
$g_{ph}= 2.4$ meV/$\text{T}^{1/2}$. }
\end{figure}

The spin and valley splittings $\Delta_S$ and $\Delta_V$
parameterize our model for the CDW phase and determine the filling
factors that enter the Hamiltonian and the optical matrix
elements. We fix $\Delta_S$ using the experimental graphene
g-factor measured in Ref.\cite{Kurganova}. We treat $\Delta_V$ as
a fitting parameter, obtaining $\Delta_V = 0.24 \times B
[\textrm{T}] \textrm{meV}$.
In our calculations, we use $g_{ph} = 2.3 \times
\sqrt{B [\textrm{T}]} \textrm{meV}$, which is in good agreement
with density functional theory (DFT) calculations \cite{Piscanec}
and experiments \cite{Lazzeri,Yan,Pisana}. We take the position of
the $E_{01}$ transition line and its broadening to be given by
$v_F=1.01 \times 10^6\textrm{ms}^{-1}$ and $\gamma_{01}
[\textrm{meV}] = 3+0.8\sqrt{B[\textrm{T}]} $ respectively,
consistent with their measured values at low magnetic fields away from the
$\Gamma$-phonon frequency. 
For the parameter $\gamma_{0}$ characterizing the broadening
of the Landau levels in Eq.\ref{EF}, we have taken
$\gamma_{0}=\gamma_{01}/2$ because the broadening of the $E_{01}$
transition should have contributions from both the $n=0$ and $n=1$
Landau levels.
The calculations for the splitting at
high magnetic fields are in excellent agreement with the
experimental transmission spectra for both samples S4 (Fig.~\ref{figS4}) and S5 (Fig.~\ref{figS5}).
We neglect $K$-phonon absorption \cite{Orlita}, which might
account for the discrepancy between theory and experiment
at the lower frequency and magnetic field range of our data.

Upon changing the carrier density $N_s$ by a factor of $\sim10$ in
our calculations, we find only minor changes in the transmission
spectra, which is consistent with the observation that samples S4
and S5 have remarkably similar transmission spectra despite having
different carrier densities. This is because the disorder induced
broadening reduces the dependence of the sublevel filling factors
($\nu_{K\uparrow}$, etc.) on the carrier density.

For symmetry breaking driven by electron-electron interactions,
the details of the screening function plays a vital role in
determining the nature of the ground state. The additional
screening afforded by the multiple graphene layers in our epitaxial
graphene samples might favor the CDW configuration, with two electrons
on the same sublattice, over the CAF state observed in hBN-supported samples
\cite{Young2}. Furthermore, coupling between rotationally misaligned
layers breaks the local A-B sublattice (i.e. valley) symmetry
\cite{He1,He2}, promoting the CDW ground state.

\section{Conclusions \label{secConc}}

In conclusion, we have used magneto-optical spectroscopy to
characterize a SU(4) symmetry broken phase in our epitaxial
graphene samples. Based on the evolution of the transmission lines
near the $\Gamma$ phonon frequency, we identify this phase as a
CDW phase for the specific samples considered, with different
occupation numbers at valleys $K$ and $K'$. Because of the
valley-sensitive nature of the electron-phonon interaction, the
transmission study used here complements spin-sensitive transport
measurements in tilted magnetic fields in the study of symmetry
breaking in graphene. Our experimental method can be applied to
open questions such as symmetry breaking of the different LLs in
graphene and bilayer graphene, as well as the effect of disorder
on the broken symmetry phase in these systems.

\section*{Acknowledgements}

L.Z.T. and the theoretical analysis were
supported by the Theory Program at the Lawrence Berkeley National
Lab through the Director, Office of Science, Office of Basic
Energy Sciences, Materials Sciences and Engineering Division, U.S.
Department of Energy under Contract No. DE-AC02-05CH11231.
Numerical simulations were supported in part by NSF Grant
DMR10-1006184. Computational resources were provided by NSF
through TeraGrid resources at NICS and by DOE at Lawrence Berkeley
National Laboratory's NERSC facility. We acknowledge the support
of this work by the France-Berkeley Fund and the European Research
Council (ERC-2012-AdG-320590-MOMB).

\noindent * Corresponding author: sglouie@berkeley.edu.

\section*{Appendix}

In general, the detailed analysis of magneto-transmission spectra requires the use
of a multi-layer dielectric model including all layer dielectric
properties of the sample. In particular, for each graphene sheet,
one has to introduce the corresponding components of the optical
conductivity tensor $\sigma_{xx}(\omega)$ and
$\sigma_{xy}(\omega)$.  Here, the $x$ and $y$-axis lie in the
plane of the sample. For instance $\sigma_{xx}(\omega)$, in a
one-electron approximation, for transitions involving the $n=0$ LL,
is written as:

\begin{equation}\label{sigmaxx}
\sigma_{xx}(\omega,B)= i
\frac{e^{3}B}{h\omega}\sum_{r,s}\frac{M^2_{r,s}(f_{r}(B)-f_{s}(B))}{\hbar\omega-E_{r,s}(B)+i\Gamma_{rs}(B)}
\end{equation}

\noindent where $r$,$s$ scan the values 0 and $\pm 1$, $0\leq
f_{r}\leq 1$ is the occupation factor of the LL $r$, $M_{r,s}$ the
optical matrix element, $E_{r,s}=E_{r}-E_{s}=E_{01}$ and
$\Gamma_{rs}(B)=\gamma_{01}(B)$ measures the broadening of the
transition. $M_{r,s}\propto v_{0}$ where $v_{0}$ is the Fermi
velocity given by LDA calculations \cite{Bychkov}. In the
present work, we have taken for all samples $v_{0}= 0.85\times
10^{6}m s^{-1}$. This is a different parameter from $v_F$ which appears in 
$E_{01}$ because the energies and wavefunctions are corrected to different extents by the electron-electron interaction \cite{Bychkov}. 
This approach requires the knowledge of the
number of effective active layers as well as their carrier
densities $N_s$ ($\nu=N_s\Phi_{0}/B$, $\Phi_{0}$ being the flux
quantum) which, in turn, implies some approximations.

The multi-layer dielectric model  assumes that each graphene sheet
is uniformly spread over the sample. This is a strong assumption,
difficult to justify \textit{a priori} and we have been lead to
correct it by assuming a mean coverage which, in the present case
for samples S4 and S5, has been determined to be about 70 per
cent. We next evaluate the number $N_{eff}$ for each sample. In
the range of magnetic fields 12 to 17 T, the relative transmission
spectra (Fig. 2, top panel) reaches values above 1 which
depends on the number $N_{eff}$: we have therefore a guide to
estimate this quantity. We estimate $N_{eff}=7$ for samples S4 and
S5.

The carrier density $N_s$ for each layer  is determined in the
following way: one knows that, for $2<\nu<6$, upon increasing $B$,
the intensity of the $E_{01}$ absorption starts to increase, at
the expense of the intensity of the $E_{12}$ transition
($E_{12}=E_{2}-E_{1}$). The intensity does not change with $B$ for
$\nu<2$. Therefore, the disappearance of the optical transition
$E_{12}$ corresponds to $\nu=2$. Following  
the transmission spectra as a function of $B$, one can evaluate the
carrier density $N_{sm}$ for each layer $m$. This is an iterative
process which converges reasonably (within 20 per cent) but has to
be done independently for each sample. The value of $N_{s1}$ for
the layer close to the SiC substrate can be set arbitrary to 5 to
6 $10^{12}cm^{-2}$ as given by transport data  on samples grown
under similar conditions: this layer indeed and the two following
ones do not contribute to the transition $E_{01}$ in the present
experiment. Finally, in the range of magnetic field larger than 27
T, where we focus our attention in this
paper, the number of optically active layers (for optical transitions
involving the $n=0$ LL) ranges between 3 to 4
for samples S4 and S5 with carrier densities ranging from 0.5 to
12 $\times 10^{11} \text{cm}^{-2}$.

\bibliography{basename of .bib file}

\begin{references}



\bibitem{Zhang} Y. Zhang, Z. Jiang, J. P. Small, M. S. Purewal,
  Y.-W. Tan, M. Fazlollahi, J. D. Chudow, J. A. Jaszczak, H. L. Stormer,
  and P. Kim,
  \textit{Phys. Rev. Lett.} \textbf{96,} 136806 (2006).

\bibitem{Zhao} Y. Zhao, P. Cadden-Zimansky, F. Ghahari, and P. Kim,
  \textit{Phys. Rev. Lett.} \textbf{108,} 106804 (2012).

\bibitem {Song} Y. J. Song, A. F. Otte, Y. Kuk, Y. Hu, D. B.
  Torrance, P. N. First, W. A. de Heer, H. Min, S. Adam, M. D. Stiles,
  A. H. MacDonald, and J. A. Stroscio,
  \textit{Nature} \textbf{467,} 185-189 (2010).

\bibitem {Miller} D. L. Miller, K. D. Kubista, G. M. Rutter, M.
  Ruan, W. A. de Heer, M. Kindermann, P. N. First, and J. A. Stroscio,
  \textit{Nat Phys} \textbf{6,} 811-817 (2010).


\bibitem {Young} A. F. Young, C. R. Dean, L. Wang, H. Ren, P.
  Cadden-Zimansky, K. Watanabe, T. Taniguchi, J. Hone, K. L. Shepard,
  and P. Kim,
  \textit{Nat Phys} \textbf{8,} 550-556 (2012).

\bibitem {Young2} A. F. Young, J. D. Sanchez-Yamagishi, B. Hunt, S.
  H. Choi, K. Watanabe, T. Taniguchi, R. C. Ashoori, and P.
  Jarillo-Herrero,
  \textit{Nature} \textbf{505,} 528-532 (2014).


\bibitem {Yu} G. L. Yu, R. Jalil, B. Belle, A. S. Mayorov, P. Blake,
  F. Schedin, S. V. Morozov, L. A. Ponomarenko, F. Chiappini, S.
  Wiedmann, U. Zeitler, M. I. Katsnelson, A. K. Geim, K. S. Novoselov,
  and D. C. Elias,
  \textit{PNAS} \textbf{110,} 3282-3286 (2013).


\bibitem {Amet} F. Amet, J. R. Williams, K. Watanabe, T. Taniguchi,
  and D. Goldhaber-Gordon,
  \textit{Phys. Rev. Lett.} \textbf{112,} 196601 (2014).


\bibitem {Kharitonov} M. Kharitonov,
  \textit{Phys. Rev. B} \textbf{85,} 155439 (2012)
  and references therein.

\bibitem {Abanin} D. A. Abanin, B. E. Feldman, A. Yacoby, and B. I.
  Halperin,
  \textit{Phys. Rev. B} \textbf{88,} 115407 (2013).

\bibitem{Sodemann} I. Sodemann and A. H. MacDonald,
  \textit{Phys. Rev.  Lett.} \textbf{112,} 126804 (2014).

\bibitem {Roy} B. Roy, M. P. Kennett, and S. D. Sarma,
  arXiv:1406.5184 (2014).

\bibitem {Henriksen} E. A. Henriksen, P. Cadden-Zimansky, Z. Jiang,
  Z. Q. Li, L.-C. Tung, M. E. Schwartz, M. Takita, Y.-J. Wang, P. Kim,
  and H. L. Stormer,
  \textit{Phys. Rev. Lett.} \textbf{104,} 067404 (2010).


\bibitem {Orlita} M. Orlita, L. Z. Tan, M. Potemski, M. Sprinkle, C.
  Berger, W. A. de Heer, S. G. Louie, and G. Martinez,
  \textit{Phys. Rev. Lett.} \textbf{108,} 247401 (2012).


\bibitem {Yang} C. H. Yang, F. M. Peeters, and W. Xu,
  \textit{Phys. Rev. B} \textbf{82,} 075401 (2010).


\bibitem {Goerbig} M. O. Goerbig, J.-N. Fuchs, K. Kechedzhi, and V.
  I. Fal'ko,
  \textit{Phys. Rev. Lett.} \textbf{99,} 087402 (2007).


\bibitem {Berger} C. Berger, Z. Song, T. Li, X. Li, A. Y. Ogbazghi,
  R. Feng, Z. Dai, A. N. Marchenkov, E. H. Conrad, P. N. First, and W.
  A. de Heer,
  \textit{J. Phys. Chem. B} \textbf{108,} 19912-19916 (2004).


\bibitem {Hass} J. Hass, F. Varchon, J. E. Mill\'{a}n-Otoya, M.
  Sprinkle, N. Sharma, W. A. de Heer, C. Berger, P. N. First, L. Magaud,
  and E. H. Conrad,
  \textit{Phys.\ Rev.\ Lett.\ } \textbf{100}, 125504 (2008).

\bibitem {Faugeras} C. Faugeras, M. Orlita, S. Deutchlander, G.
  Martinez, P. Y. Yu, A. Riedel, R. Hey, and K. J. Friedland,
  \textit{Phys. Rev. B} \textbf{80,} 073303 (2009).


\bibitem {Sadowski} M. L. Sadowski, G. Martinez, M. Potemski, C.
  Berger, and W. A. de Heer,
  \textit{Phys. Rev. Lett.} \textbf{97,} 266405 (2006).


\bibitem {Toyozawa1}  Y. Toyozawa, M. Inoue, T. Inui, M. Okazaki,
  and E. Hanamura,
  \textit{ J.\ Phys.\ Soc.\ Jpn.\,} \textbf{22} 1337-1349 (1967).

\bibitem {Kurganova} E. V. Kurganova, H. J. van Elferen, A.
  McCollam, L. A. Ponomarenko, K. S. Novoselov, A. Veligura, B. J. van
  Wees, J. C. Maan, and U. Zeitler,
  \textit{Phys. Rev. B} \textbf{84}, 121407(R) (2011).

\bibitem {Piscanec} S. Piscanec, M. Lazzeri, F. Mauri, A. C.
  Ferrari, and J. Robertson,
  \textit{Phys. Rev. Lett.} \textbf{93,} 185503 (2004).


\bibitem {Lazzeri} M. Lazzeri, C. Attaccalite, L. Wirtz, and F.
  Mauri,
  \textit{Phys. Rev. B} \textbf{78,} 081406 (2008).


\bibitem {Yan} J. Yan, Y. Zhang, P. Kim, and A. Pinczuk,
  \textit{Phys. Rev. Lett.} \textbf{98,} 166802 (2007).


\bibitem {Pisana} S. Pisana, M. Lazzeri, C. Casiraghi, K. S.
  Novoselov, A. K. Geim, A. C. Ferrari, and F. Mauri,
  \textit{Nat Mater} \textbf{6,} 198 - 201 (2007).


\bibitem{Fuchs} J.-N. Fuchs and P. Lederer,
  \textit{Phys. Rev. Lett.} \textbf{98,} 016803 (2007).

\bibitem{Jung} J. Jung and A. H. MacDonald,
  \textit{Phys. Rev. B} \textbf{80,} 235417 (2009).

\bibitem{He1} L. Meng, Z.-D. Chu, Y. Zhang, J.-Y. Yang,
  R.-F. Dou, J.-C. Nie, and L. He,
  \textit{Phys. Rev. B} \textbf{85,} 235453 (2012).

\bibitem{He2} J.-B. Qiao and L. He,
  \textit{Phys. Rev. B} \textbf{90,} 075410 (2014).

\bibitem {Bychkov}  Yu.\ A.\ Bychkov and G.\ Martinez,
 Phys.\ Rev.\ B  \textbf{77}, 125417 (2008).

\end{references}

\end{document}